\documentstyle[12pt,epsfig]{article}
\def\plb#1{Phys.~Lett.~{\bf B#1}}
\def\npb#1{Nucl.~Phys.~{\bf B#1}}
\def\prl#1{Phys.~Rev.~Lett.~{\bf #1}}
\def\prd#1{Phys.~Rev.~{\bf D#1}}

\def\e{\epsilon}

\def\L{\lambda}

\def\e3{$\epsilon_3$}

\def\bsgam{$b\rightarrow s\gamma$ }
\def\ch2{$\chi^2$}

\def\co#1{{\ifmmode{\cal O}_{#1}\else${\cal O}_{#1}$\fi}}

\def\dltmh{$\Delta m_H^2\;$}
\def\dltmb{$\Delta m_b \;$}
\def\mupos{$\mu > 0 \;\;$}

\newdimen\unit
\def\point#1 #2 #3{\vbox to0pt{\kern-#2\unit
  \hbox{\kern#1\unit#3}\vss}
 \nointerlineskip}

\newcommand{\be}{\begin{equation}}
\newcommand{\ee}{\end{equation}}
\newcommand{\bea}{\begin{eqnarray}}
\newcommand{\eea}{\end{eqnarray}}

\newcommand{\gev}{\mbox{ GeV }}

\begin{document}
\thispagestyle{empty}
\noindent
\begin{flushright}
        CERN-TH/2001-176 \\
                July 2001 
\end{flushright}

\vspace{1cm}
\begin{center}
  \begin{Large}
  \begin{bf}
Predictions for Higgs and SUSY spectra from SO(10)
Yukawa Unification with \mupos

\end{bf}
 \end{Large}
\end{center}
  \vspace{1cm}
    \begin{center}
T. Bla\v{z}ek$^\dagger$, R. Derm\' \i \v sek$^*$ and S. Raby$^*$\\
      \vspace{0.3cm}
\begin{it}
$^\dagger$Department of Physics, 
University of Southampton,
Southampton, UK \\
                   and Faculty of Mathematics and Physics,
           Comenius Univ., Bratislava, Slovakia\\
$^*$Department of Physics, The Ohio State University,
174 W. 18th Ave., Columbus, Ohio  43210
\end{it}
  \end{center}
  \vspace{1cm}
\centerline{\bf Abstract}
\begin{quotation}
\noindent
 We use $t,\; b,\;\tau$ Yukawa unification to
constrain SUSY parameter space.  We find a narrow region survives for 
$\mu > 0$ (suggested by \bsgam and the anomalous magnetic moment of the
muon) with  $A_0 \sim - 1.9 \; m_{16}$, $m_{10} \sim 1.4\; m_{16}$,
 $m_{16} \sim 1200 -3000$ \gev 
and $\mu, M_{1/2} \sim 100 - 500$ \gev.   Demanding Yukawa unification thus 
makes definite predictions for Higgs and sparticle masses.
\end{quotation}

\newpage

Minimal supersymmetric [SUSY] SO(10) 
grand unified theories [GUTs] have many profound features~\cite{so10}:
all fermions in one family sit in one {\bf 16} dimensional
spinor representation; the two Higgs doublets of the minimal 
supersymmetric standard model sit in one {\bf 10} dimensional fundamental representation, and gauge coupling unification at a GUT 
scale $M_G \sim 3 \times 10^{16}$ \gev fits well with the low energy 
data~\cite{susygut,gutexp}.
In addition in the simplest version of SO(10) the third generation Yukawa couplings  are given by a single term in the superpotential 
$W =  \L \; {\bf 16 \; 10 \; 16}$ resulting in Yukawa unification  $\L_t = \L_b = \L_\tau = \L_{\nu_\tau} \equiv \; {\bf \L}$ and a prediction for   $M_t$
  with large $\tan\beta \sim 50$~\cite{nothreshcorr}.   
\footnote{Note, GUT scale threshold corrections to this Yukawa unification boundary condition are naturally small ( $<$ 1\% ),
since they only come at one loop from the SO(10) gauge sector and the third generation - Higgs Yukawa coupling~\cite{wright}.  This is in
contrast to GUT scale threshold corrections to gauge coupling 
unification which  may be significant, coming from doublet/triplet
splitting in the Higgs sector and, even more importantly, the SO(10) breaking sector which typically has many degrees of freedom.
The data requires \e3 $= \frac{\alpha_3(M_G) - \alpha_G(M_G)}{ \alpha_G(M_G)} \sim$ -4\%~\cite{chi2}.}
This beautiful result is however marred by potentially
large weak scale threshold corrections~\cite{threshcorr,pierceetal}
$$ m_b(M_Z) = 
\L_b(M_Z) \; \frac{v}{\sqrt{2}} \; cos\beta \; (1 + \Delta m_b^{\tilde g} +  \Delta m_b^{\tilde \chi^+} + \Delta m_b^{\tilde \chi^0} + \Delta m_b^{\log})  .$$  
For \mupos the gluino term is positive and in most regions of SUSY parameter
space it is the dominant contribution to \dltmb.   Reasonable fits
prefer \dltmb $< 0$; hence Yukawa unification is easy to satisfy with
$\mu < 0$.

The decay \bsgam and the muon anomalous magnetic moment also get
significant corrections proportional to $\tan\beta$\cite{threshcorr}.
The SUSY contribution to \bsgam comes from one loop diagrams similar to those contributing
to the bottom mass.  
The chargino term typically dominates and has opposite sign to the SM and charged Higgs contributions, thus reducing the branching ratio for $\mu > 0$.  This is necessary to fit the data
since the SM contribution is somewhat too big.  $\mu < 0$ would on the other
hand constructively add to the branching ratio and is problematic.
In addition, the recent measurement of the anomalous magnetic moment of the muon
$a^{NEW}_\mu = (g - 2)/2 = 43 \; (16) \times 10^{-10}$ also favors \mupos~\cite{muon}.
Thus it is important to confirm that Yukawa unification can work consistently with \mupos.

In this paper we assume exact Yukawa unification and search, using a $\chi^2$
analysis, for regions of SUSY parameter space with
\mupos providing good fits to the low energy data.  
We show that Yukawa unification dramatically constrains the Higgs and SUSY spectra.  These results are sensitive to the SUSY breaking
mechanism. 

It is much easier to obtain EWSB with large $\tan\beta$ when the
  Higgs up/down masses are split ($m_{H_u}^2 < m_{H_d}^2$)~\cite{ewsb}.
In our analysis we consider two particular 
schemes we refer to as
universal and D term splitting.  
In the first case the third generation 
squark and slepton soft masses are given by the universal mass parameter $m_{16}$ , and only Higgs masses are split:  $m_{(H_u, \; H_d)}^2 =
m_{10}^2 \;( 1 \mp \Delta m_H^2)$.  
In the second case we assume D term splitting, i.e. that the D term 
for $U(1)_X$ is non-zero,
where $U(1)_X$ is obtained in the decomposition of $SO(10) \rightarrow
SU(5) \times U(1)_X$.  In this second case, we have
$  m_{(H_u,\; H_d)}^2 =  m_{10}^2 \mp 2 D_X $, $
 m_{(Q,\; \bar u,\; \bar e)}^2 =  m_{16}^2 + D_X$,  $m_{(\bar d,\; L)}^2 =  m_{16}^2 - 3 D_X $.
The universal case does not at 
first sight appear to be similarly well motivated.  It is quite 
clear however that in any SUSY model the
Higgs bosons are very special.  R parity is used to distinguish
Higgs superfields from quarks and leptons.   In addition, a 
supersymmetric mass term $\mu$ with value of order the weak scale is 
needed for an acceptable low energy phenomenology.   
Since $\mu$ is naturally of order $M_G$, one needs some symmetry argument 
why it is suppressed.  Of course, if the Higgs are special, then 
it is reasonable to assume splitting of Higgs,
while maintaining universal squark and slepton masses.  This
may be achieved by GUT scale threshold corrections to the soft
SUSY breaking scalar masses~\cite{ewsb}.  Here we present the most
compelling mechanism~\cite{bdr}.  In $SO(10)$, 
neutrinos necessarily have a Yukawa term 
coupling active neutrinos to the ``sterile" neutrinos present in the {\bf 16}.
In fact for $\nu_\tau$ we have $\L_{\nu_\tau} \; \bar \nu_\tau \; L \; H_u$ 
with $\L_{\nu_\tau} = \L_t = \L_b = \L_\tau \equiv \; {\bf \L}$.   
In order to obtain a tau neutrino with mass $m_{\nu_\tau} \sim 0.05$ eV (consistent with atmospheric neutrino oscillations), the ``sterile" $\bar \nu_\tau$ must obtain  a Majorana mass $M_{\bar \nu_\tau} \geq 10^{13}$ GeV.   Moreover, since neutrinos couple to $H_u$ (and not to $H_d$) with a fairly large Yukawa coupling (of order 0.7), they naturally distinguish the two Higgs multiplets.  With $\L = 0.7$ and $M_{\bar \nu_\tau} = 10^{13}$ GeV, we obtain a 
significant GUT scale threshold correction with $\Delta m_H^2 \approx 10$\%, remarkably close to the value needed to fit the data.   At the same time, we obtain a 
small threshold correction to Yukawa unification $\approx 2.5$\% (for more
details see \cite{bdr}).

\begin{figure}[t]
\begin{center}

\input{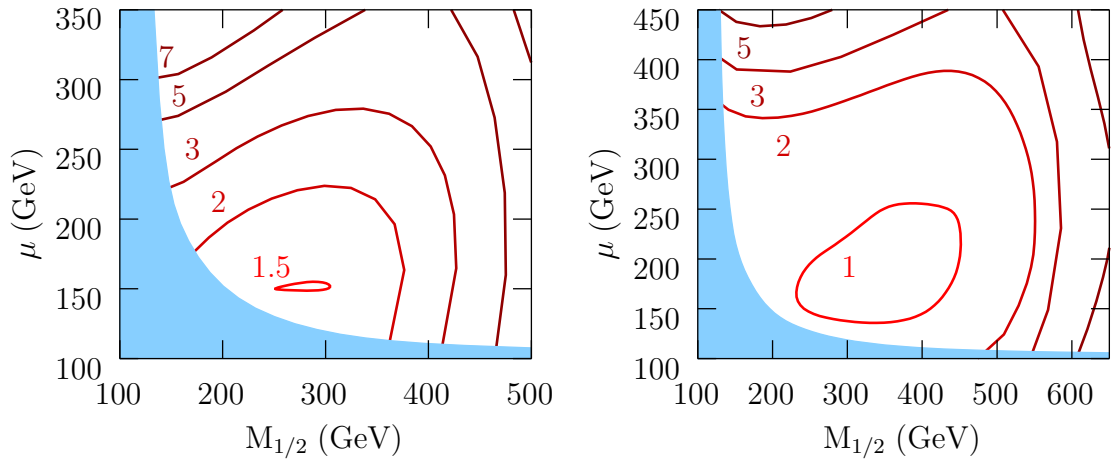}

\caption{$\chi^2$ contours for $m_{16} = 1500$\gev (Left) and $m_{16} = 2000$
\gev (Right).  The shaded region is excluded by the chargino mass limit
$m_{\tilde \chi^+} > 103$ GeV.}
\label{figure:chi2}
\end{center}
\end{figure}

Our analysis is a top-down approach with 11 input parameters, defined
at $M_G$, varied to minimize a $\chi^2$ function composed
of 9 low energy observables.
The 11 input parameters are: 
$M_G, \; \alpha_G(M_G),$ $ \epsilon_3$; the Yukawa coupling
$\L$, and the 7 soft SUSY breaking parameters
$\mu,\; M_{1/2},\; A_0, \; \tan\beta$, $m_{16}^2, \; m_{10}^2$,  \dltmh  
 ($D_X $) for universal (D term) case. 
We use two (one)loop renormalization group [RG] running 
for dimensionless (dimensionful) parameters from $M_G$ to
$M_Z$ and complete one loop threshold corrections at $M_Z$~\cite{pierceetal}.
  We require electroweak symmetry breaking using an improved Higgs potential, including $m_t^4$ and $m_b^4$ corrections in an effective 2 
Higgs doublet model below $M_{stop}$~\cite{carenaetal}. 
   The \ch2 function includes the 9 observables; 6 precision electroweak data $\alpha_{EM},\; 
G_\mu, \;  \alpha_s(M_Z),\; M_Z, \; M_W, \; \rho_{NEW}$ and the 3 fermion 
masses $M_{top},\;  m_b(m_b), \; M_\tau$.  The experimental values used
for the low energy observables are given in the table.

\begin{figure}[t]
\begin{center}

\input{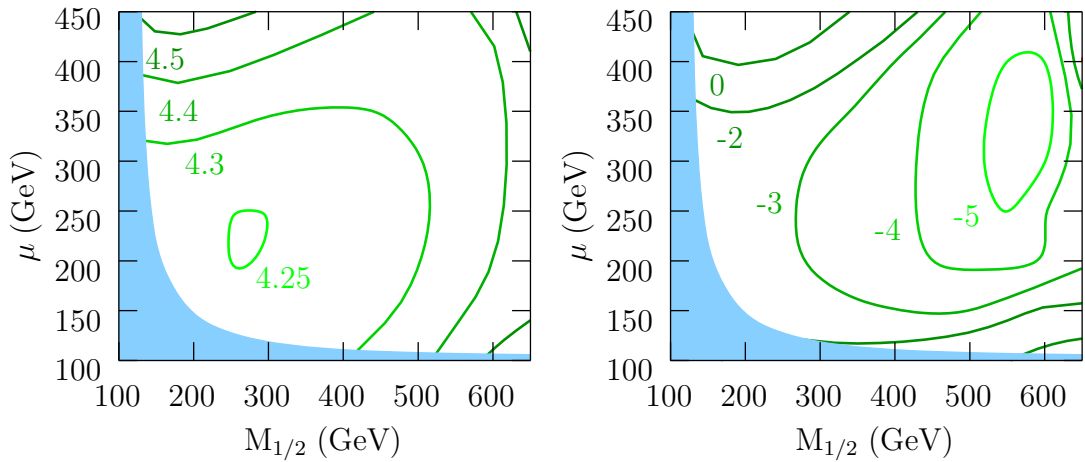}

\caption{Contours of constant $m_b(m_b)$[GeV] (Left) and \dltmb in \% (Right)
for $m_{16} = 2000$\gev.}
\label{figure:m_b_and_corr}
\end{center}
\end{figure}

Fig. 1 shows the constant $\chi^2$ contours for $m_{16} = 1500$ and $2000$ \gev
in the case of universal squark and slepton masses.
We find acceptable fits ($\chi^2 < 3$) for $A_0 \sim - 1.9 \; m_{16}$, $m_{10} \sim 1.4 \; m_{16}$ and $ m_{16} \geq 1.2$ TeV.  The
best fit is for $m_{16} \geq 2000$ \gev with $\chi^2 < 1$.  
Note, electroweak symmetry
breaking in this region of parameter space requires splitting  Higgs up/down masses, \dltmh $\sim $ O(13\%).  This range of soft SUSY parameters
is consistent with solution (B) of Olechowski 
and Pokorski~\cite{ewsb}.   
In the table we present the input parameters, the fits and the
predicted Higgs and SUSY spectra for
two representative points with universal squark and slepton masses
and the best fit value for D term splitting.   We have not presented
the contour plots for D term splitting since as can be seen from
the best fit point in the table, the bottom quark mass is 
poorly fit in this case and $\chi^2 > 5$.  Recall, since we 
have 11 input parameters and
only 9 observables, we consider such poor fits unacceptable.

Fig. 2 gives the constant $m_b(m_b)$ and \dltmb contours for $m_{16} = 2000$ \gev.
We see that the best fits, near its central value, are found
with \dltmb $\leq - 2$\%.   
Why does Yukawa unification only work in this narrow region of SUSY
parameter space? 
The log corrections $\Delta m_b^{\log} \sim 4 - 6$\% (total contribution
from gluino, neutralino, chargino and electroweak loops)  are positive and
they must be cancelled in order to obtain \dltmb $\leq - 2$ \%. 
The leading mass insertion corrections proportional to 
$\tan\beta$ are approximately given by~\cite{threshcorr}
$$ \Delta m_b^{\tilde g} \approx  \frac{2 \alpha_3}{3 \pi} \;
\frac{\mu m_{\tilde g}}{m_{\tilde b}^2} \; tan\beta  \hspace{.4in} {\rm and} \hspace{.4in}
 \Delta m_b^{\tilde \chi^+} \approx \frac{\L_t^2}{16 \pi^2} \; 
\frac{\mu A_t}{m_{\tilde t}^2} \; tan\beta.  $$
 They can
naturally be as large as 40\%.
The chargino contribution is typically opposite in sign to the gluino, 
since $A_t$ runs to an infrared fixed point $\propto - M_{1/2}$(see for example, Carena 
et al.~\cite{threshcorr}).  Hence in order to cancel the positive
contribution of both the log and gluino contributions, a large negative 
chargino contribution is needed.   This can be accomplished for 
$- A_t > m_{\tilde g}$ and  $m_{\tilde t_1} << m_{\tilde b_1}$.  The first 
condition can be satisfied for $A_0$ large and negative, which helps pull 
$A_t$ away from its infrared fixed point.   The second condition is also 
aided by large $A_t$.  However in order to obtain a large enough splitting between $m_{\tilde t_1}$ and $m_{\tilde b_1}$, large values of $m_{16}$ are needed.    Note, that for universal scalar masses, the lightest stop is typically lighter than the sbottom.  We typically find $m_{\tilde b_1} \sim
3 \; m_{\tilde t_1}$.
On the other hand, D term splitting with 
$D_X > 0$ gives $m_{\tilde b_1} \leq m_{\tilde t_1}$.   As a result in the case 
of universal boundary conditions excellent fits are obtained for top, 
bottom and tau masses; while for D term splitting the best fits give 
$m_b(m_b) \geq 4.59$\gev.   

\begin{figure}[t]
\begin{center}

\input{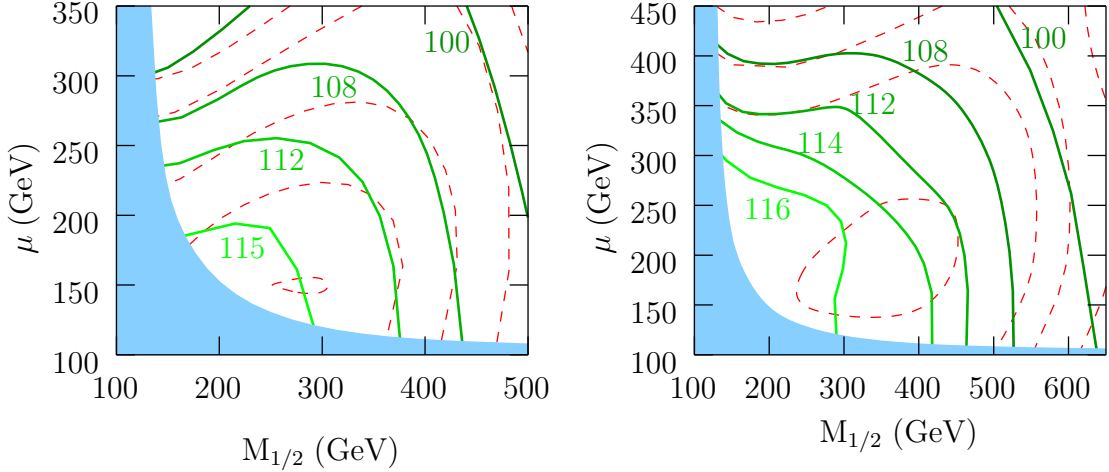}

\caption{Contours of constant $m_h$ [GeV] (solid lines) with $\chi^2$ contours from Fig. 1 (dotted lines) for $m_{16} = 1500$\gev (Left) and
$m_{16} = 2000$\gev (Right).}
\label{figure:h0_and_chi2}
\end{center}
\end{figure}

\protect
\begin{table}
\caption[8]{
{\bf Three representative points of the fits. } \\
We fit the central values: $M_Z = 91.188,\; M_W = 80.419,\; 
G_{\mu}\times 10^5  = 1.1664,\; \alpha_{EM}^{-1} = 137.04,\;
M_{\tau} = 1.7770$ with 
0.1\% numerical uncertainties; and the following with the experimental
uncertainty in parentheses:
$\alpha_s(M_Z) = 0.1180\; (0.0020), \;
\rho_{new}\times 10^3 =  -0.200\; (1.1), $ $M_t  = 174.3\; (5.1), \;
m_b(m_b)  = 4.20\; (0.20)$.
The neutral Higgs masses $h,\; H, \; A_0$ are pole masses; while all other
sparticle masses are running masses.

}
\label{t:table}
$$
\begin{array}{|l|c|c|c|}
\hline
{\rm Data \; points}  &  1 &  2 & 3  \\
\hline
 {\rm Input\; parameters}    &   &     &  \\
\hline
\;\;\;\alpha_{G}^{-1}  & 24.46 &  24.66      & 24.73  \\
\;\;\; M_G \times 10^{-16}& 3.36  &  3.07     &  3.13 \\
\;\;\;\epsilon_3 &  -0.042&  -.0397     & -0.046  \\
\;\;\;\L  & 0.70  & 0.67    & 0.80  \\
\hline
\;\;\; m_{16}  & 1500 &  2000   &  2000 \\
\;\;\; m_{10}  & 2027 &  2706  &  2400 \\
\;\;\; \Delta m_H^2  &   0.13 &  0.13 &  0.07 \\
\;\;\; M_{1/2}  & 250 &  350  &  350 \\
\;\;\;\mu  & 150 & 200  &  115 \\
\;\;\; \tan\beta  & 51.2  &  50.5  &   54.3 \\
\;\;\; A_0   & - 2748 &  -3748   & - 731  \\
\hline
\hline
\chi^2 \; {\rm observables}   &    &   &\\
\hline
\;\;\;M_Z  & 91.13 &  91.14 &     91.15    \\
\;\;\;M_W  & 80.45&  80.45 &   80.44   \\
\;\;\;G_{\mu}\times 10^5  & 1.166 &  1.166 &  1.166  \\
\;\;\;\alpha_{EM}^{-1} & 137.0 &  137.0   &  137.0 \\
\;\;\;\alpha_s(M_Z) & 0.1175 &  0.1176  &   0.1161 \\
\;\;\;\rho_{new}\times 10^3 & 0.696& 0.460  & 0.035 \\
\hline
\;\;\;M_t       & 175.5 &  174.6    & 177.9 \\
\;\;\;m_b(m_b)   &  4.28  &    4.27 & 4.59     \\
\;\;\;M_{\tau}   & 1.777 &  1.777   &  1.777 \\
\hline
 {\rm TOTAL}\;\;\;\; \chi^2  & 1.50 & 0.87  & 5.42  \\
\hline
\hline
\;\;\; h_0              &  116  & 116 &  115 \\
\;\;\; H_0               &  120  & 121 & 117 \\
\;\;\; A_0               &  110  & 110 &  110 \\
\;\;\; H^+               &  148  & 148 &  146 \\
\;\;\; \tilde \chi^0_1   &  86 & 130 & 86 \\
\;\;\; \tilde \chi^0_2   &  135  & 190 & 126 \\
\;\;\; \tilde \chi^+_1   &   123 & 178 & 105  \\ 
\;\;\; \tilde g           &   661 &  913 & 902 \\
\;\;\; \tilde t_1         &  135  & 222 & 1020 \\
\;\;\; \tilde b_1        &  433 & 588  & 879  \\
\;\;\; \tilde \tau_1      & 288   & 420 & 1173 \\
\hline
\;\;\; a_\mu^{SUSY} \times 10^{10}      &  9.7 & 5.5 & 6.1 \\
\hline
\end{array}
$$
\end{table}

Finally in Fig. 3 we show the constant light Higgs mass contours
 for $m_{16} = 1500$ and $2000$ \gev 
(solid lines)
with the constant $\chi^2$ contours overlayed (dotted 
lines).  Yukawa unification for 
$\chi^2 \leq 1$  clearly prefers light Higgs mass in a narrow range,  112 -  118\gev.   In this region
the CP odd, the heavy CP even Higgs and the 
charged Higgs bosons are also quite light (see fit 2 in the table).\footnote{
It would be interesting to see how sensitive our results, for
Higgs masses, may
be to alternative electroweak symmetry breaking approximations.  In this
paper we have used the effective 2 Higgs doublet analysis of \cite{carenaetal}
with an estimated 3 \gev uncertainty in Higgs masses.
This approximation may be particularly well suited to the light Higgs
spectrum we obtain in our analysis.  The alternative scheme, in which
the Higgs tadpoles are evaluated at a scale of order $M_{stop}$
~\cite{pierceetal} is however more frequently used in the literature.}  
In addition we find the mass of $\tilde t_1 \sim (150 - 250)$ GeV, $\tilde b_1 \sim (450 - 650)$ GeV, $\tilde \tau_1 \sim (200 - 500)$ GeV, $\tilde g \sim (600 - 1200)$ GeV, $\tilde \chi^+ \sim (100 - 250)$ GeV, and $\tilde \chi^0 \sim (80 - 170)$ GeV.
All first and second generation squarks and sleptons have mass of
order $m_{16}$.  The light stop  and chargino may be visible at
the Tevatron.  With this spectrum we expect $\tilde t_1 \rightarrow 
\tilde \chi^+ \;b $ with
$\tilde \chi^+ \rightarrow \tilde \chi^0_1 \; \bar l \; \nu$ to be dominant.
Lastly $\chi^0_1$ is the LSP and possibly a good dark matter candidate~\cite{leszek}.

The region of SUSY parameter space preferred by Yukawa unification
may be consistent with a supergravity mechanism for 
SUSY breaking at $M_{Pl}$ with RG running from $M_{Pl}$ to $M_G$
(see for example Murayama et al.~\cite{ewsb}).  
It however cannot be obtained
with gauge mediated or gaugino mediated SUSY breaking mechanisms
where $A_0 = 0$ at zeroth order.
It may also be obtained in anomally mediated schemes but in this case
one still has to worry about slepton masses squared and also the fact
that in this case, since the gluino and chargino masses have opposite
sign, it is difficult to fit both \bsgam and $a_\mu$.   

In a future paper~\cite{bdr} we present the sparticle spectrum in more 
detail and consequences for Tevatron searches.
We discuss the sensitivity of our results to small GUT scale threshold corrections to Yukawa unification with both universal and D term Higgs up/down splitting.  We also check the robustness of the Higgs spectrum by
artificially adjusting the CP odd Higgs mass using a penalty in 
$\chi^2$.   We find that $\chi^2$ increases by at most 40\% for any $m_{A^0}$ less than
$ \approx 350$ GeV.    The light Higgs mass $m_h$ is rather insensitive to the 
value of $m_{A^0}$; whereas $m_H, m_{H^+}$ are linearly dependent on $m_{A^0}$.
We also consider constraints resulting from the
processes $b \rightarrow s \gamma$, $B_s \rightarrow \mu^+ \; \mu^-$, $a^{NEW}_\mu$ and the proton lifetime in a semi model independent way.  
We shall only make a few short comments here.  In order to fit $b \rightarrow s \gamma$ we find that the coefficient $C_7^{MSSM} \sim - C_7^{SM}$
(see for example, Eqn. 9 in Ref. \cite{br}) with the chargino term dominating
by a factor of order 5 over all other contributions.   This is due to the
light stop $\tilde t_1$.   In fact, $b \rightarrow s \gamma$ is more sensitive
to $m_{\tilde t_1}$ than $m_b(m_b)$.  Fitting the central value
$B(b \rightarrow s \gamma) = 2.96 \times 10^{-4}$~\cite{gambinomisiak}
requires a heavier $\tilde t_1$ with
$(m_{\tilde t_1})_{MIN} \sim 500$ GeV; significantly larger than the range which
provides the best fits to $m_b$. We now find $m_b(m_b)_{MIN} \sim 4.3$.  Moreover no other sparticle masses are affected.
The process $B_s \rightarrow \mu^+ \; \mu^-$ provides a lower bound on $m_{A^0} \geq 200$ GeV (see recent work of ~\cite{bsmumu}).~\footnote{We thank
K.S. Babu and C. Kolda for discussions.}   However this has
only a minor impact on $\chi^2$ as discussed above.   We recall that 
proton decay
experiments prefer values of $m_{16} > 2000$ GeV and $m_{16} >> M_{1/2}$ (see ref. ~\cite{pdecay}).   This is in accord with the range of SUSY
parameters found consistent with third generation Yukawa unification.
There is however one experimental result which is not consistent with
either Yukawa unification or proton decay and that is the anomalous magnetic
moment of the muon.  Large values of $m_{16} \geq 1200$\gev lead to very small values for  $a^{NEW}_\mu \leq 16 \times 10^{-10}$.  Hence a necessary
condition for Yukawa unification is that forthcoming BNL data~\cite{muon} and/or a reanalysis of the strong interaction contributions to $a_\mu^{SM}$ will significantly
decrease the discrepancy between the data and the standard model
value of $a_\mu$.   

In summary, most of the results of our analysis
including only third generation fermions remain intact when incorporating
flavor mixing.  The light Higgs mass and most sparticle masses receive only
small corrections.  The lightest stop mass increases, due to 
$b \rightarrow s \gamma$.  Nevertheless there is still a significant
$\tilde t_1 - \tilde t_2$ splitting and $m_{\tilde t_1} << 
m_{\tilde b_1}$.  The $A^0, \; H, \; H^+$ masses are necessarily
larger in order to be consistent with $B_s \rightarrow \mu^+ \; \mu^-$~\cite{bdr}, which suggests that this process should be observed soon;
possibly at Run II of the Tevatron.
Finally, the central value for $a_\mu^{NEW}$ must significantly decrease.
The ``smoking guns" of SO(10) Yukawa unification, presented in this 
letter, should be observable at Run III of the Tevatron or at LHC.  Also,
in less than a year we should have more information on $a_\mu^{NEW}$.

In previous works Yukawa unification with \mupos was not possible.\footnote{
In ref. ~\cite{chi2} the corrections $\Delta m_b^{\log}$ were neglected and Yukawa unification with \mupos was obtained
in a large region of parameter space with no constraints on the Higgs and
SUSY spectra.}   Pierce et al.~\cite{pierceetal} assume  \dltmh $= 0$ and, as a result, they are not able to enter the region of SUSY parameter space consistent
with both EWSB and Yukawa unification.
Baer et al.~\cite{baeretal} also cannot obtain Yukawa unification
with \mupos.   This is because they use D term splitting for Higgs up/down
which as discussed typically leads to sbottom lighter than stop.

While completing this article, the paper by Baer and Ferrandis~\cite{baerferrandis}  appeared which confirmed our results~\cite{susy01}
on the existence of a preferred region of SUSY parameter space consistent
with Yukawa unification and \mupos.  Their results however
require significant GUT threshold corrections to $\L_t = \L_b = \L_\tau$ of order 8 - 15\% which helps them obtain $m_{\tilde t_1} < m_{\tilde b_1}$.    They also claim that better fits are obtained with
D term splitting than with the universal splitting case.  We believe the latter
 is only
true because the authors do not allow their SUSY parameters,
in particular $m_{16}$ and $A_0$, to explore the region of parameter space discussed in ~\cite{susy01} and this paper.~\footnote{In version 3 of this
paper, the authors now cover a larger region of soft SUSY breaking parameter
space.   Now solutions with D term splitting are only found with $\geq 28$\%
GUT threshold corrections to Yukawa unification.  Moreover, they still claim
that D term splitting works better than the universal case.  We have no
explanation for this discrepancy.}

{\bf Acknowledgements}

S.R. and R.D. are partially supported by DOE grant DOE/ER/01545-816.
S.R. graciously acknowledges the Alexander von Humboldt Foundation, the Universit\"at Bonn and the Theory Division at CERN for partial support.
R.D. and S.R. thank the Physics Department at the Universit\"at Bonn and 
the Theory Division at CERN for their kind hospitality while working on this
project.   R.D. and S.R. thank A. Dedes for his help and collaboration
in the early stages of this analysis.

%

\end{document}